\documentclass[pre,preprint,showkeys,amsfonts,amssymb,amsmath,eqsecnum,letterpaper,nofootinbib]{revtex4}

\usepackage{graphicx}

\begin{document}

\renewcommand{\r}{\mathbf{r}}
\newcommand{\ren}{{\text{ren}}}
\newcommand{\hr}{{\hat{r}}}
\newcommand{\ha}{{\hat{a}}}
\newlength{\GraphicsWidth}
\setlength{\GraphicsWidth}{11cm}

\title{Guest charges in an electrolyte: renormalized charge, long- and
short-distance behavior of the electric potential and density
profiles}

\author{Gabriel T\'ellez}
\email{gtellez@uniandes.edu.co} 
\affiliation{Departamento de F\'{\i}sica, Universidad de Los Andes,
A.A.~4976, Bogot\'a, Colombia}

\begin{abstract}
We complement a recent exact study by L.~\v{S}amaj on the properties
of a guest charge $Q$ immersed in a two-dimensional electrolyte with
charges $+1/-1$. In particular, we are interested in the behavior of
the density profiles and electric potential created by the charge and
the electrolyte, and in the determination of the renormalized charge
which is obtained from the long-distance asymptotics of the electric
potential. In \v{S}amaj's previous work, exact results for arbitrary
coulombic coupling $\beta$ were obtained for a system where all the
charges are points, provided $\beta Q<2$ and $\beta < 2$. Here, we
first focus on the mean field situation which we believe describes
correctly the limit $\beta\to 0$ but $\beta Q$ large. In this limit we
can study the case when the guest charge is a hard disk and its charge
is above the collapse value $\beta Q>2$. We compare our results for
the renormalized charge with the exact predictions and we test on a
solid ground some conjectures of the previous study. Our study shows
that the exact formulas obtained by \v{S}amaj for the renormalized
charge are not valid for $\beta Q>2$, contrary to a hypothesis put
forward by \v{S}amaj. We also determine the short-distance asymptotics
of the density profiles of the coions and counterions near the guest
charge, for arbitrary coulombic coupling. We show that the coion
density profile exhibit a change of behavior if the guest charge
becomes large enough ($\beta Q\geq 2-\beta$). This is interpreted as a
first step of the counterion condensation (for large coulombic
coupling), the second step taking place at the usual Manning--Oosawa
threshold $\beta Q=2$.
\end{abstract}


\maketitle

\section{Introduction}
\label{sec:Intro}

In a recent paper~\cite{Samaj-guest-charges}, \v{S}amaj studied the
properties of one or two ``guest'' charges immersed in a classical
(i.e.~non-quantum) two-dimensional two-component charge-symmetric
electrolyte. Using results from the (1+1)-integrable sine-Gordon
model~\cite{Zamalod2, Destri-Vega,Zamolod-conformal-norm}, in
particular the known expressions for the expectation value of the
exponential field~\cite{Lukyanov-Zamolod-exp-field, Fateev-reflexion}
and for the form factors~\cite{Lukyanov-form-fac1, Lukyanov-form-fac2}
of this theory, and the exact solution for the thermodynamics of the
two-dimensional two-component plasma~\cite{Samaj-Travenec-TCP}, he
was able to determine exactly the excess chemical potential of a
single ``guest'' charge immersed in the electrolyte, the long-distance
behavior of the electric potential created by this guest charge and
the long-distance behavior of the interaction energy between two guest
charges, in the whole regime where the system of point charges is
stable (i.~e.~when both the guest charges and the internal charges of
the electrolytes are point particles).

An important result from Ref.~\cite{Samaj-guest-charges} is for the
electric potential created by a single guest charge $Q$ immersed in
the electrolyte. This potential $\psi(r)$ has a long-distance
behavior, as the distance $r\to\infty$, similar to the screened
potential predicted by Debye--H\"uckel theory,
\begin{equation}
  \label{eq:psi-long-distance}
  \psi(r)\sim Q_{\ren} K_0(m_1 r)
\end{equation}
where $m_1$ is the inverse screening length (it is also the mass of
the lightest breather of the sine-Gordon model), and it is given in
terms of the inverse Debye length $\kappa=\sqrt{2\pi\beta n}$ in
equation~(4.15) of Ref.~\cite{Samaj-guest-charges}. We shall use the
same notations as in Ref.~\cite{Samaj-guest-charges}: $\beta$ is the
Coulombic coupling, the electrolytes charges are $+1/-1$ and $n$ is
the density. In~(\ref{eq:psi-long-distance}), $K_0$ is the modified
Bessel function of order 0. However $Q_{\ren}$ is not the charge $Q$
of the guest charge (as it would be in Debye--H\"uckel theory), but it
is known as the renormalized charge. \v{S}amaj found the following
expression for the renormalized charge~(equations (5.8) and (5.9) of
Ref.~\cite{Samaj-guest-charges})
\begin{equation}
  \label{eq:renorm-charge}
  Q_{\ren}=\frac{2 \exp\left[-\int_{0}^{\pi\beta/(4-\beta)}
      \frac{t\,dt}{\pi\sin t}\right] }{(4-\beta)
    \sin\left(\frac{\pi\beta}{2(4-\beta)}\right)} \,
  \sin\left(\frac{\pi\beta Q}{4-\beta}\right) \,.
\end{equation}
The concept of renormalized charge is very important in colloidal
science~\cite{Alexander,Trizac-Bocquet-Aubouy-PRL, Belloni, Hansen,
Levin, Tellez-Trizac-charge-sat, Tellez-Trizac-PB-large-a}, thus
\v{S}amaj result is of extreme importance for colloidal science, in
particular for the study of cylindrical polyelectrolytes, which can be
reduced to a two-dimensional problem.

\v{S}amaj claims that the rigorous validity of his
result~(\ref{eq:renorm-charge}) is for $\beta|Q|<2$, which is the
regime where the system of point particles is stable. However, he
gives some arguments to support a conjecture he called
``regularization hypothesis''. This conjecture states that the
validity of~(\ref{eq:renorm-charge}) goes actually beyond $\beta |Q|
=2$. In the case $\beta |Q|>2$, the regularization hypothesis says
that equation~(\ref{eq:renorm-charge}) gives the value of the
renormalized charge for a guest particle of charge $Q$ and radius $a$
in the limit $m_1 a \ll 1$.

In this article, we present some indications that suggest that the
regularization hypothesis is not valid. These indications come from
the small-coupling limit $\beta\to 0$ but when $\beta |Q|$ can be
arbitrary large. This will be explained in Section~\ref{sec:PB}.

In Section~\ref{sec:restrictions}, follows a discussion on the
short-distance behavior of the density profiles, near the guest
charge. In particular, we show that the coion density have a change of
behavior when $\beta |Q| = 2-\beta$, which can be interpreted as a
``precursor'' of the counterion condensation.


\section{The mean field limit: Poisson--Boltzmann equation}
\label{sec:PB}

\subsection{The case $a=0$ and $\beta |Q|<2$}

Let $\psi(r)$ be the electric potential at a distance $r$ from a
single guest charge $Q$ immersed in the electrolyte. The guest charge
is an impenetrable disk of radius $a$ with its charge spread over its
perimeter. We shall use the dimensionless potential $y(\hr)=\beta
\psi(r)$ with $\hr=\kappa r$. 

For a three dimensional electrolyte in the presence of an arbitrary
external charge distribution, it is rigorously proved in
Ref.~\cite{Kennedy-MFT} that, in the limit $\beta\to 0$, the density
and correlation functions of the electrolyte are given by the ones of
an ideal gas in the presence of the external field $y(\hr)$ which is
the solution of the nonlinear Poisson--Boltzmann equation with the
external source charge. Based on this evidence, we conjecture that
this is also valid for our problem, although in our case we consider a
two-dimensional system, and in the case $a\neq 0$ we include a
hard-core interaction between the external guest charge and the
electrolyte (which is not considered in the proof of
Ref.~\cite{Kennedy-MFT}). Thus, assuming the validity of this
hypothesis, the mean field electric potential $y(\hr)$ for our problem
is the solution of
\begin{eqnarray}
  \label{eq:PB}
  \Delta_{\mathbf{\hr}} y(\hr) = \sinh(y(\hr)) && \quad\hr>\ha\\
  \label{eq:Laplace}
  \Delta_{\mathbf{\hr}} y(\hr) = 0  &&\quad \hr<\ha
\end{eqnarray}
satisfying the boundary conditions
\begin{eqnarray}
  \label{eq:BC-infty}
  \lim_{\hr\to\infty} y(\hr)&=&0\\
  \label{eq:BC-a}
  \lim_{\hr\to \ha^{+}} \hr \frac{dy(\hr)}{\hr}&=&-\beta Q\\
  \label{eq:continuity}
  \lim_{\hr\to\ha^{-}} y(\hr)&=&\lim_{\hr\to\ha^{+}} y(\hr)
\end{eqnarray}
where $\ha=\kappa a$ is the guest particle radius in units of
$1/\kappa$ and the charge $Q$ of the guest particle is supposed to be
uniformly spread over its perimeter.

This is usually called the mean field approximation. Let us remind the
reader that the mean field approximation corresponds to the classical
treatment of the sine-Gordon model: Poisson--Boltzmann equation is the
stationary action equation of the sine-Gordon model. Let us also
clarify, that the nonlinear Poisson--Boltzmann theory is correct in
the limit $\beta\to 0$ when it is used to describe the density
profiles of the electrolyte created by an \textit{external} charge
distribution, under the conditions considered in
Ref.~\cite{Kennedy-MFT}. On the other hand, the linear Debye--H\"uckel
theory should be used to describe the distribution functions of the
\textit{internal} charges of the bulk electrolyte in the limit
$\beta\to 0$~\cite{Kennedy-DH}.

The two-dimensional Poisson--Boltzmann equation~(\ref{eq:PB}) has been
solved exactly~\cite{McCoy-Tracy-Wu,McCaskill-Fackerell,
Tracy-Widom-polyelec}, and in particular the connexion problem has
been studied extensively. This is the problem to relate the
long-distance behavior of $y(\hr)$ with its short-distance
behavior~\cite{McCaskill-Fackerell,
Tracy-Widom-Toda-asympt,Tellez-Trizac-PB-cyl-exact}.  This connexion
problem is essential for the determination of the renormalized charge.

Let us recall some of the results
from~\cite{McCoy-Tracy-Wu,McCaskill-Fackerell, Tracy-Widom-polyelec,
Tellez-Trizac-PB-cyl-exact} relevant for our discussion. For a point
guest particle, $a=0$, and $\beta|Q| <2$, the electric potential has
the long-distance behavior
\begin{equation}
  \label{eq:large-asymptotics}
  y(\hr) \sim 4 \lambda K_0(\hr)\ , \quad \hr\to\infty
\end{equation}
and the short-distance behavior
\begin{equation}
  \label{eq:small-asymptotics}
  y_{11}(\hr)=-2A\ln\hr+2\ln B + o(1)
  \,, \quad \hr\to 0\,.
\end{equation}
The solution of the connexion problem
\cite{McCoy-Tracy-Wu,McCaskill-Fackerell,Tracy-Widom-Toda-asympt}
states that the constant $\lambda$ intervening in the long-distance
behavior is related to the constants $A$ and $B$ of the short-distance
behavior by
\begin{equation}
  A=\frac{2}{\pi}\arcsin (\pi\lambda)
\end{equation}
and 
\begin{equation}
  B= 2^{3A} \frac{\Gamma\left(\frac{1+A}{2}\right)}{
    \Gamma\left(\frac{1-A}{2}\right)}
\end{equation}
where $\Gamma()$ is the Gamma function. In relation to the physical
problem, one immediately recognizes that $2A=\beta Q$ is the bare
charge of the guest particle and $4\lambda=\beta Q_{\ren}$ is the
renormalized charge. Thus, the renormalized charge is given by
\begin{equation}
  \label{eq:Qren-PB}
  \beta Q_{\ren}=\frac{4}{\pi} \sin\left(\frac{\pi \beta Q}{4}\right)
  \,.
\end{equation}
 Notice that~(\ref{eq:Qren-PB}) corresponds to the limit $\beta\to 0$
 while keeping $\beta Q$ arbitrary large (with $\beta|Q|<2$) of
 equation~(\ref{eq:renorm-charge}). Indeed the mean field
 approximation for the study of single guest charge $Q$ immersed in the
 electrolyte is asymptotically correct in the low coupling limit
 $\beta\to0$ and $\beta Q$ arbitrary. 

\subsection{The case $a\neq 0$ and $\beta |Q|>2$}

If $\beta |Q|>2$ it is mandatory to consider that the guest particle
is a disk with impenetrable radius $a\neq 0$, otherwise the charges of
opposite sign of the electrolyte will collapse with the guest
charge. Although it is not possible (yet) to obtain exact results when
$a\neq 0$ for arbitrary values of $\beta$, in the limit $\beta\to 0$
we can obtain some results, under the hypothesis that the mean field
approach is correct in that limit when $a\neq 0$. Thus, when $a\neq
0$, and $\beta\to 0$ and $\beta |Q|>2$, we assume that the electric
potential is given again by the mean field theory: it is the solution
of equations~(\ref{eq:PB}) and~(\ref{eq:Laplace}) and the boundary
conditions~(\ref{eq:BC-infty}), (\ref{eq:BC-a})
and~(\ref{eq:continuity}). We will consider the case $\kappa a=\ha \ll
1$ but $a\neq 0$.

To formally solve this problem, let us introduce $y_0(\hr)$ the
solution of Poisson-Boltzmann equation~(\ref{eq:PB}) \textit{in the
whole space}, for $\hr\in\mathbb{R}^{+}$, and satisfying the boundary
condition~(\ref{eq:large-asymptotics}) when $\hr\to\infty$. So far,
$\lambda$ in equation~(\ref{eq:large-asymptotics}) can be seen as an
integration constant. Then, the electric potential is given by
\begin{equation}
  y(\hr)=
  \begin{cases}
    y_0(\ha)&\text{for\ }\hr\leq\ha\\
    y_0(\hr)&\text{for\ }\hr>\ha
  \end{cases}
\end{equation}
Enforcing the boundary condition~(\ref{eq:BC-a}) at $\hr=\ha$
determines the integration constant $\lambda$. 

To determine $\lambda$ in the case $\ha\ll1$ we need to know the
short-distance asymptotics of $y_0(\hr)$.  Without loss of generality
(because the electrolyte is charge symmetric), let us suppose that
$Q>0$.  For $\beta Q<2 +O(1/|\ln \ha|)$\footnote{When $a\neq 0$ there
is a small (negative) correction of order $1/|\ln a|$ to the critical
value $\beta Q=2$, for details
see~\cite{Trizac-Tellez-lettre-Manning}.}, the short-distance
asymptotics are the same as in the previous section, given in
equation~(\ref{eq:small-asymptotics}).

If $\beta Q$ is large enough (larger than 2) then $\lambda >1/\pi$,
and then the short-distance behavior of $y_0(\hr)$ changes
drastically. It is now given by~\cite{McCoy-Tracy-Wu,
McCaskill-Fackerell}
\begin{equation}
\label{eq:asympt-above-y}
y_0(\hr)=-2\ln\left(\frac{-\hr}{4\mu}\right)
-2\ln\left[\sin\left(2\mu\ln\frac{\hr}{8}+2\phi(\mu)\right)\right]
+O(\hr^4)\ , \quad \hr\to 0
\end{equation}
with
\begin{equation}
\label{eq:asympt-above11-phi-Gamma}
\phi(\mu)=\arg (\Gamma(1-i\mu))
\,.
\end{equation}
Let us define
\begin{equation}
  \varphi(\hr,\mu)=2\mu\ln\frac{\hr}{8}+2\phi(\mu) 
\end{equation}
which is the argument of the sine function
in~(\ref{eq:asympt-above-y}). The constant $\mu$ appearing in the
short-distance behavior of $y_0(\hr)$ is related to $\lambda$~(see
equation~(\ref{eq:asympt-above11-mu}) below) and can be determined by
the boundary condition~(\ref{eq:BC-a}) at $r=a$. It is the solution of
the transcendental equation
\begin{equation}
  \label{eq:mu-Q}
  \beta Q=2 + 4\mu \cot[2\mu\ln\frac{\ha}{8}+2\phi(\mu)]
  \,.
\end{equation}

The solution of the connexion problem gives the following relation
between $\lambda$ and $\mu$~\cite{McCoy-Tracy-Wu, McCaskill-Fackerell}
\begin{equation}
\label{eq:asympt-above11-mu}
\mu=-\frac{1}{\pi}\cosh^{-1}(\pi\lambda)
\end{equation}
Notice that we choose here $\mu<0$. If $\beta Q>2$ the argument of the
$\cot$ function in equation~(\ref{eq:mu-Q}) is in the range
$[\pi/2,\pi]$. Equation~(\ref{eq:asympt-above11-mu}) allows us to
obtain the renormalized charge $Q_{\ren}$.

For $a\neq 0$, the definition of the renormalized charge comes from
the long-distance behavior of the electric potential and its
comparison with the solution from the linear Debye--H\"uckel theory
\begin{equation}
  y(\hr)\sim \frac{\beta Q_{\ren}}{\ha K_1(\ha)}\, K_0(\hr)
  \,,
  \quad \hr\to\infty
\end{equation}
where $K_1$ is the modified Bessel function of order 1. Comparing
with~(\ref{eq:large-asymptotics}) we have $\beta Q_{\ren}=4 \ha
K_1(\ha) \lambda$. Notice the additional factor $\ha K_1(\ha)$ in the
renormalized charge. However this factor is not really important since
if $\ha\ll 1$, $\ha K_1(\ha)\sim 1$. We shall see that it is the
behavior of $\lambda$ which will be very different from the case when
$a=0$ and $\beta Q<2$. Once $\mu$ has been determined from
equation~(\ref{eq:mu-Q}), using equation~(\ref{eq:asympt-above11-mu})
we can find the renormalized charge
\begin{equation}
  \label{eq:Qren-PB-above}
  \beta Q_{\ren}= \frac{4 \ha K_1(\ha)}{\pi}\,  \cosh(\pi \mu)
\end{equation}

Let us comment a few points on the small-$\hr$ behavior of the
potential $y(\hr)$ and of $y_0(\hr)$. From
equation~(\ref{eq:asympt-above-y}), one can distinguish three special
regions. Notice that $\varphi(\hr,\mu)$ is a decreasing function of
$\hr$ since $\mu<0$. The first region is for $\hr=0$ up to a value
$r^{*}$ such that
\begin{equation}
  \label{eq:def-r*}
  \varphi(r^{*},\mu)=\pi\,.
\end{equation}
In this region, the formal solution $y_0(\hr)$ of Poisson--Boltzmann
equation has no physical meaning, since $\varphi(\hr,\mu)$ decreases
from $+\infty$ to $\pi$, and then the argument of the logarithm of the
second term of equation~(\ref{eq:asympt-above-y}) oscillates around
zero, changing of sign, thus $y(\hr)$ is not always real but can
become complex. However this region is inside the guest charge
($r^{*}\leq\ha $) and in this region the electric potential is a
constant: $y(\hr)=y_0(\ha)$.

The second region is for $\hr\in[\ha,r_M]$ where $r_M$ is given by
\begin{equation}
  \varphi(r_M,\mu)=\pi/2 \,.
\end{equation}
In this region $\pi/2<\varphi(\hr,\mu)<\pi$. If $\hr$ is close to
$\ha$ (and $\ha$ is close to $r^{*}$), the second term of $y(\hr)$ in
equation~(\ref{eq:asympt-above-y}) can be very large because
$\varphi(\hr,\mu)$ is close to $\pi$. Then, as $\hr$ increases,
$\varphi(\hr,\mu)$ decreases from $\pi$ down to $\pi/2$, and then, the
second term of $y(\hr)$ in equation~(\ref{eq:asympt-above-y})
decreases very fast. Since the counterion density is proportional to
$e^{y(\hr)}$, this indicates that close to the guest charge there is a
very large counterion density, which decreases quite fast as the
distance $\hr$ increases. This is a manifestation of a widely known
phenomenon in the theory of cylindrical polyelectrolytes, known as the
Manning--Oosawa counterion condensation~\cite{Manning,Oosawa}. The
layer of ``condensed'' counterions extends from $\hr=\ha$ to
$\hr=r_M$.

When $\hr=r_M$ the second term of equation~(\ref{eq:asympt-above-y})
vanishes. At this point the total charge
$Q_{\text{guest}+\text{condens}}$ of the guest particle plus the
condensed counterion layer is such that $\beta
Q_{\text{guest}+\text{condens}}=2$, as it can be seen from the first
term of equation~(\ref{eq:asympt-above-y}). We recover here another
characteristic of the Manning--Oosawa counterion
condensation~\cite{Manning,Oosawa,Kholodenko-condens-PIII}: the layer
of condensed counterions reduces the bare charge of the guest charge
to $\beta Q_{\text{guest}+\text{condens}}=2$. Above $r_M$, we enter a
third region, outside the condensed layer of counterions, where $\hr$
starts to become large enough such that the small-$\hr$
behavior~(\ref{eq:asympt-above-y}) is no longer valid.

From a physical argument we can now see that as $\beta Q$ becomes
larger than 2, the renormalized charge is not given anymore by
equation~(\ref{eq:Qren-PB}) (which is valid only for $a=0$ and $\beta
Q<2$). Indeed, from the discussion above, we know that in the region
$r^{*}>r_M$, just outside the counterion condensed layer, the guest
charge ``dressed'' with the condensed counterions can be seen as an
object with a charge $Q_{\text{guest}+\text{condens}}=2/\beta$. Thus
its renormalized charge will be close to the prediction of
equation~(\ref{eq:Qren-PB}) for $\beta Q=2$, that is $\beta Q_{\ren}$
is close to $4/\pi$.

The bare charge could be arbitrary large (with $\beta Q >2$), but the
renormalized charge will remain close to $4/\pi$, because of the
Manning--Oosawa counterion condensation phenomenon. In particular, the
renormalized will not oscillate and become eventually negative as
predicted by~(\ref{eq:Qren-PB}) for $\beta Q>2$, if the regularization
hypothesis was valid.

This argument can also be justified from a more rigorous point of
view. If $\ha$ is small enough, the solution of
equation~(\ref{eq:mu-Q}) for $\mu$ is small (of order $1/|\ln
(\ha/8)|$). The renormalized charge is given by the correct
formula~(\ref{eq:Qren-PB-above}) when $\beta Q>2$. If $\mu\ll 1$ and
$\ha \ll 1$, we see from~(\ref{eq:Qren-PB-above}) that $\beta
Q_{\ren}$ is close to $4/\pi$. It is actually slightly larger than
$4/\pi$.

\begin{figure}
  \includegraphics[width=\GraphicsWidth]{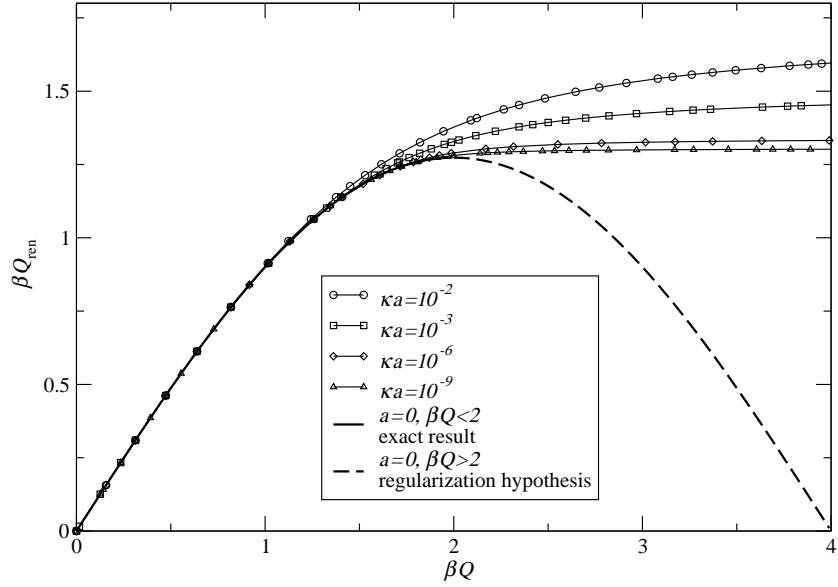}
  \caption{
    \label{fig:Qren}
    The renormalized charge $Q_{\ren}$ as a function of the bare
    charge $Q$, in the mean-field limit $\beta\to 0$, for various
    values of the radius $a$ of the guest charge. For $a=0$, the exact
    result~(\ref{eq:Qren-PB}) is shown in full thick line for $\beta
    Q<2$.  In dashed line, the extension of~(\ref{eq:Qren-PB}) for
    $\beta Q>2$ is shown: this is the prediction of the regularization
    hypothesis from Ref.~\cite{Samaj-guest-charges}. The symbols
    correspond to values $a>0$, obtained from a numerical resolution
    of Poisson--Boltzmann equation.  }
\end{figure}

This is verified numerically in figure~\ref{fig:Qren}, where we
computed the renormalized charge from a numerical resolution of
Poisson--Boltzmann equation, using the method described in
Ref.~\cite{Trizac-Langmuir}. We confirm numerically that the
renormalized charge, for $\beta Q>2$, is slightly above $4/\pi$ (for
$\ha \ll 1$). This can be contrasted with the prediction of the
regularization hypothesis, shown in dashed line in
figure~\ref{fig:Qren}, where $Q_{\ren}$ is expected to decay when
$\beta Q>2$ and even vanish and change its sign at $\beta Q=4$. The
numerical results show that this is not true: $Q_{\ren}$ is always an
increasing function of $Q$, and eventually saturates to a finite value
for large values of $Q$.

%
%
%
%
%

This saturation phenomenon~\cite{Kholodenko-condens-PIII} of the
renormalized charge is quite usual in the nonlinear Poisson--Boltzmann
approach to the problem. When the saturation phenomenon occurs we also
have $\ha=r^{*}$. Indeed, if $\ha=r^{*}$, by the
definition~(\ref{eq:def-r*}) of $r^{*}$ we have $\varphi(\ha,\mu)=\pi$
and one can verify that in equation~(\ref{eq:mu-Q}),
$Q\to+\infty$. Solving
\begin{equation}
  \label{eq:sat-mu}
  \varphi(\ha,\mu)=\pi
\end{equation}
for $\mu$ and replacing in~(\ref{eq:Qren-PB-above}), allow us to
obtain the saturation value of the renormalized
charge. Equation~(\ref{eq:sat-mu}) is a transcendental equation, but
since $\mu$ is small, of order $1/|\ln(\ha/8)|$, if $\ha\ll 1$, it can
be solved in an expansion of powers of $1/|\ln(\ha/8)|$. For example,
up to order 4 in $1/|\ln(\ha/8)|$, we find the renormalized charge at
saturation~\cite{Tellez-Trizac-PB-cyl-exact}
\begin{equation}
  \label{eq:xi-eff-sat-11}
  \beta Q_{\ren}^{\text{sat}}= \frac{4 \ha
    K_1(\ha)}{\pi}\left[\cosh\frac{\pi^2}{2\left(\ln\frac{\ha}{8}+
    \gamma\right)}+O\left(|\ln\ha|^{-5}\right)\right]
\end{equation}
where $\gamma\simeq 0.5772$ is the Euler constant. 

In conclusion to this part, we notice the failure of the
regularization hypothesis in the limit $\beta\to 0$ with $\beta Q>2$
and $\ha\ll 1$. The renormalized charge is not given by
equation~(\ref{eq:renorm-charge}) in that limit as the regularization
hypothesis claims.


\section{Short-distance behavior of the density profiles: A ``precursor'' of the counterion condensation at $\beta Q
  =2 -\beta$}
\label{sec:restrictions}

An important factor, which is responsible of the failure of the
regularization hypothesis exposed in the previous section, is the
change of behavior of the electric potential at short distances when
$\beta Q>2$. In this section we consider the general situation when
$0<\beta<2$ and we return to the case of point particles $a=0$.  We
study the short-distance behavior of the density profiles and show
that there is a change of behavior in the asymptotic expansion at
short distances of the coion density profiles when $\beta
|Q|=2-\beta$.

The short-distance behavior of the density profiles, in the presence
of the guest charge, can be obtained by adapting an argument presented
in Refs.~\cite{Hansen-Viot,Samaj-asym} for the correlation functions.
Let suppose, without loss of generality, that $Q>0$. From the general
principles of statistical mechanics we know that the short-distance
behavior of the density profiles, near the guest charge at the origin,
is dominated by the Boltzmann factor of the Coulomb potential
$e^{-\beta Q q v_c(r)}$, with $v_c(r)=-\ln r$ the Coulomb
potential. We have
\begin{equation}
  \label{eq:n(r)-short}
  n_q(r)\sim n_q c_{Qq}\, r^{\beta Q q}\,,\quad r\to 0
\end{equation}
with $q=\pm 1$, $n_{q}$ is the bulk density of charges $q$, and the
constant $c_{Qq}$ is related to the excess chemical potentials
($\mu^{\text{exc}}_q$, $\mu^{\text{exc}}_Q$, $\mu^{\text{exc}}_{Q+q}$)
of the charges $q$, $Q$ and $Q+q$, which can in turn be expressed as
expectation values of exponentials of the sine-Gordon field $\phi$,
\begin{equation}
  \label{eq:c}
  c_{Qq}=\exp[-\beta (
  \mu^{\text{exc}}_{Q+q} - \mu^{\text{exc}}_Q + - \mu^{\text{exc}}_Q
  )]=\frac{\langle e^{ib (Q+q) \phi}\rangle}{\langle e^{ib Q \phi} \rangle
  \langle e^{ib q \phi}\rangle }
\end{equation}
with the same conventions as in~\cite{Samaj-guest-charges} for the
normalization of the sine-Gordon field and $b^2=\beta/4$. In the case
$q=+1$ (coions, same sign as $Q$) this is valid provided $\beta Q$ is
small enough, as we will explain below.

Let $\Xi[Q]$ be the grand canonical partition function of the system
composed by the electrolyte and the guest charge $Q$ fixed at the
origin, with fugacities $z_+$ and $z_-$ for positive and negative
particles respectively.  The partition function is well defined for
point particles if $\beta <2$ and $\beta |Q|<2$. The density of
particles of charge $q$ is
\begin{equation}
  \label{eq:n(r)-general}
  n_q(\r)=z_q r^{\beta Q q} \,\frac{\Xi[Q;q,\r]}{\Xi[Q]}
\end{equation}
where $\Xi[Q;q,\r]$ is the partition function of the electrolyte in
the electric field created by a guest charge $Q$ fixed at the origin
and a charge $q$ fixed at $\r$.

When $r\to 0$, $\Xi[Q;q,\r]$ has a finite limit provided that $\beta
|Q+q| <2$, since $\Xi[Q;q,0]=\Xi[Q+q]$ is the partition function of a
system composed by the electrolyte and a guest charge $Q+q$ at the
origin. Under this condition we can affirm that the short-distance
behavior~(\ref{eq:n(r)-short}) for the density profile is valid. This
can also be seen from~(\ref{eq:c}): the expectation value $\langle
e^{ib (Q+q) \phi}\rangle$ is finite provided $\beta
|Q+q|<2$~\cite{Samaj-guest-charges}.

For $Q>0$ and $q=1$, the above condition reads $\beta Q
<2-\beta$. When $\beta Q > 2-\beta $, equation~(\ref{eq:n(r)-short})
is no longer valid. We must return to the general
expression~(\ref{eq:n(r)-general}), and study the short-distance
behavior of $\Xi[Q;q,\r]$. If $\beta Q > 2-\beta $, $\Xi[Q;q,\r]$
diverges as $r\to 0$. Its short-distance behavior is dominated by the
approach of a charge $-1$ to the system of the charges $Q$ and $+1$
which are separated by a distance $r$~\cite{Hansen-Viot}. This gives
\begin{equation}
  \Xi[Q;q,\r]\sim \text{cst}\times r^{-\beta (Q+1)+2}\,,
  \quad r\to 0
\end{equation}
with $\text{cst}$ some constant, which will not be needed for our
analysis, but that can eventually be evaluated~\cite{Dotsenko-Fateev1,
Dotsenko-Fateev2}.  Thus, the short-distance behavior of the coion
density profile, for $\beta Q> 2-\beta$, is
\begin{equation}
  \label{eq:n+-above}
  n_{+}(r)\sim \text{cst} \times r^{2-\beta}
  \,,
  \quad r\to 0
\end{equation}
Notice that, on the other hand, the counterion density profile
$n_{-}(r)$ behaves always as predicted by
equation~(\ref{eq:n(r)-short}), since the corresponding $\Xi[Q;-1,\r]$
has always a finite limit at $r=0$, provided $\beta < 2$ and $\beta Q
<2$.

The mean interaction potential $w_{+,Q}(r)$ of a coion (charge +1)
with the guest charge $Q$ and its polarization cloud, is defined by
$n_{+}(r)=n_+ e^{-\beta w_{+,Q}(r)}$. From~(\ref{eq:n+-above}), we
deduce that its short-distance behavior is
\begin{equation}
  \label{eq:w-short}
  \beta w_{+,Q}(r)=
  \begin{cases}
    \beta Q \ln r \,+\, O(1)\,,& \text{if\ }\beta Q < 2-\beta\\
    (2-\beta)\ln r \,+\, O(1)\,, & \text{if\ } \beta Q > 2-\beta
  \end{cases}
  \,, \quad r\to 0
\end{equation}
 Notice that, in particular, the short-distance leading behavior of
 $w_{+,Q}(r)$ is independent of $Q$ when $\beta Q > 2-\beta$. This
 situation can be interpreted as a ``precursor'' of the
 Manning--Oosawa counterion condensation. When $\beta Q$ increases
 above $2-\beta$, the counterion cloud near the guest charge reduces
 its bare charge so that the coions ``see'' a ``dressed'' object with
 charge $2/\beta-1$, which is independent of $Q$.

On the other hand, the counterion mean interaction potential has
always the behavior (provided $\beta <2$ and $\beta Q<2$)
\begin{equation}
  \beta w_{-,Q}(r) \sim  \beta Q \ln r\,+\,O(1)
  \,, \quad r\to 0
  \,.
\end{equation}
The counterions continue to ``see'' the bare guest charge $Q$ even if
$\beta Q > 2-\beta$.

When $\beta Q\geq 2$, we arrive at the collapse of the charge $Q$ with
the counterions. In this case we need to consider that the guest
charge is an impenetrable disk with radius $a\neq 0$. $\beta Q=2$
corresponds to the well-known Manning threshold for counterion
condensation.

We would like to stress that the above analysis is valid for large
coupling $0<\beta<2$. The usual presentation of the counterion
condensation phenomenon~\cite{Manning, Oosawa} is done in a
small-coupling approximation $\beta\to 0$. Notice that in this case
both limits $\beta Q < 2-\beta $ and $\beta Q< 2$ coincide. Here we
have put in evidence that, at large coulombic coupling $\beta$, the
counterion condensation might actually take place in two steps: first
when $\beta Q = 2-\beta$, where the coion density changes its
short-distance behavior, but not the counterion density, and a second
step, at the usual threshold $\beta Q = 2$.

Let us conclude this section with a conjecture, suggested by
\v{S}amaj~\cite{Samaj-private-comm}. The mean field analysis of the
previous section shows that the formula~(\ref{eq:renorm-charge}) for
the renormalized charge is not valid beyond $\beta Q=2$ in the limit
$\beta\to 0$. For arbitrary values of $\beta$, the validity
of~(\ref{eq:renorm-charge}) might be beyond $\beta Q=2$. In the case
$\beta\to 0$, the failure of~(\ref{eq:renorm-charge}) when $\beta Q>2$
is accompanied by a change of behavior in the short distance
asymptotics of the electric potential.

For arbitrary $\beta$, when $\beta Q>2$ it is necessary to introduce a
hard core for the guest particle. The coion density will certainly
change its short distance behavior at $\beta Q=2$. On the other hand
there are indications that the counterion density will still behave at
short distances as $r^{-\beta Q}$ up to $\beta Q< 2+\beta$. This is
because, as previously noted, $\Xi[Q;-1,\r]$ remains finite up to
$\beta Q< 2+\beta$, for small $\r$. Since the counterion density
determines the dominant behavior at short distances of the electric
potential, the electric potential might actually not change its short
distance behavior until $\beta Q>2+\beta$. If this is true, the
regularization hypothesis put forward by \v{S}amaj, might be valid up
to $\beta Q< 2+\beta$~\cite{Samaj-private-comm}. Interestingly, this
leaves open the possibility of charge inversion (i.e.~the renormalized
charge becomes negative) if $\beta>1$, when $\beta Q>4-\beta$.


\section{Conclusion}

As a complement to Ref.~\cite{Samaj-guest-charges}, we have studied
the low-coupling, mean field situation, $\beta \to 0$, but $\beta |Q|$
arbitrary, in order to determine the behavior of the renormalized
charge when $\beta |Q| > 2$ of a guest charge $Q$ immersed in an
electrolyte. We have shown that, at least in this mean field
situation, the regularization hypothesis put forward by \v{S}amaj in
Ref.~\cite{Samaj-guest-charges} is not valid: the
formula~(\ref{eq:renorm-charge}) for the renormalized charge is not
valid when $\beta |Q| > 2$.

We have also studied the short-distance asymptotics of the density
profiles. The coion density profile exhibits a change of behavior if
the guest charge $Q$ is large, $\beta Q> 2 -\beta$, as shown in
equation~(\ref{eq:w-short}). Colloquially speaking, it is like if the
coions ``see'' a ``dressed'' charge $2/\beta-1$ instead of $Q$, when
$\beta Q>2-\beta$. We interpret this as a first step in the
Manning--Oosawa counterion condensation when the coulombic coupling
$\beta$ is large, the second step taking place when $\beta Q=2$ as it
is usually explained in the literature~\cite{Manning,Oosawa} for the
small coupling $\beta\to 0$ situation.

\section*{Acknowledgments}

I thank L.~\v{S}amaj for a careful reading of this manuscript and for
his comments and remarks. I also thank B.~Jancovici for his comments.

This work was partially funded by COLCIENCIAS (project 1204-05-13625),
ECOS-Nord/COLCIENCIAS, and Comit\'e de Investigaciones de la Facultad
de Ciencias de la Universidad de los Andes.


%
\newpage
\hbox{}
\vfill
\begin{center}
\includegraphics[angle=90,height=16cm]{Qren}
\end{center}
\vfill
\begin{center}
G.~T\'ellez \hskip 5cm Figure \ref{fig:Qren}
\end{center}
%
%

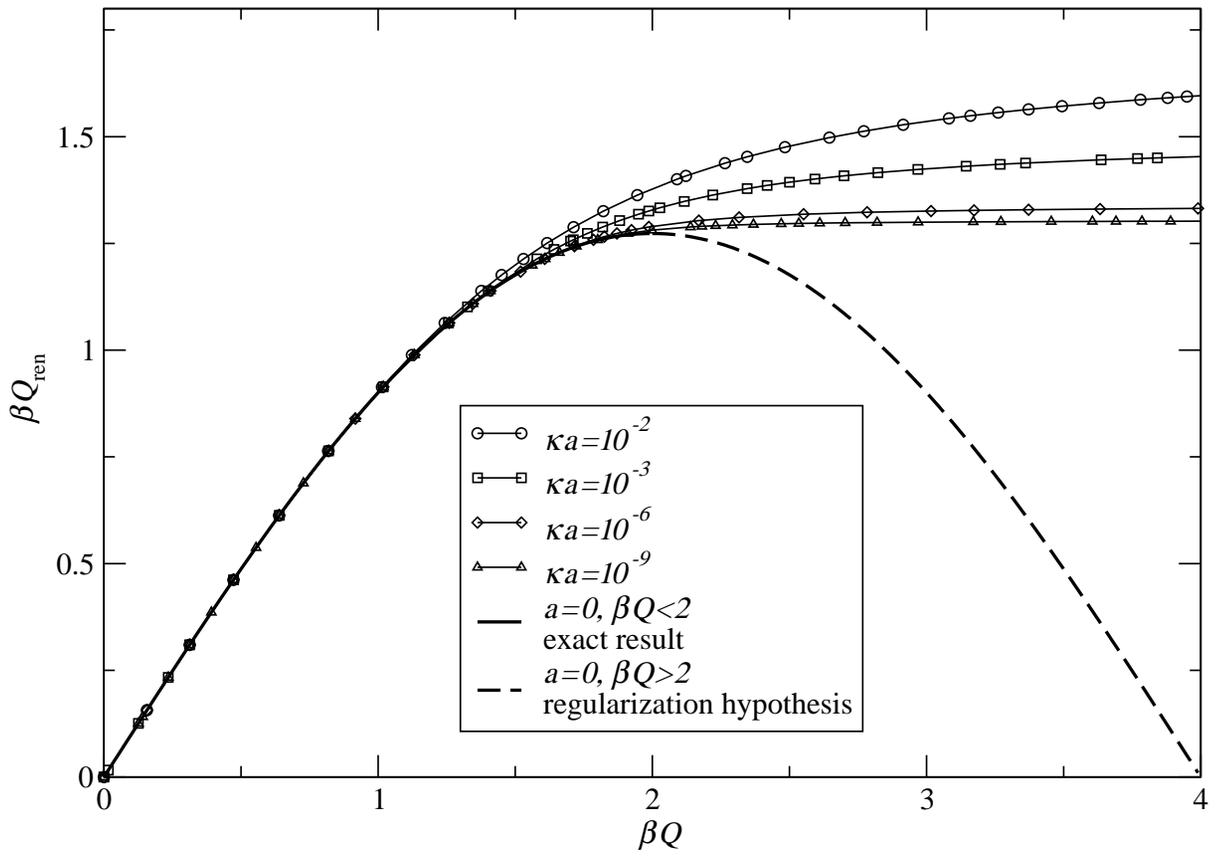
\end{document}